\documentclass[journal,twoside,web]{ieeecolor}
\usepackage{generic}
\usepackage{cite}
\usepackage[colorlinks, linkcolor=black,
anchorcolor=blue, urlcolor=cyan
]{hyperref}
\usepackage{amsmath,amssymb,amsfonts}
\usepackage{algorithmic}
\usepackage{graphicx}
\def\BibTeX{{\rm B\kern-.05em{\sc i\kern-.025em b}\kern-.08em
    T\kern-.1667em\lower.7ex\hbox{E}\kern-.125emX}}
\begin{document}
\title{Cryogenic Modeling of MOSFET Device Based on BSIM and EKV Models}
\author{Tengteng Lu, Yuanke Zhang, Yujing Zhang, Jun Xu, Guoping Guo and Chao Luo
    \thanks{Manuscript received XX XX, 2021; revised XX XX, 2021. This work was supported in part by the National Key Research and Development Program of China (Grant No. 2016YFA0301700), in part by the National Natural Science Foundation of China (Grants No. 11625419). (Corresponding author: Chao Luo)}        
    \thanks{T Lu is with the Department of Physics, University of Science and Technology of China, Hefei, Anhui 230026, China.}
    \thanks{Y Zhang, Y Zhang are with the School of Microelectronics from University of Science and Technology of China, Hefei, Anhui 230026, China.}
    \thanks{J Xu is with the Department of Physics, University of Science and Technology of China, Hefei, Anhui 230026, China.}
    \thanks{G Guo is with the Key Laboratory of Quantum Information, University of Science and Technology of China, Hefei, Anhui 230026, China.}    
    \thanks{C Luo is with the Key Laboratory of Quantum Information, University of Science and Technology of China, Hefei, Anhui 230026, China(e-mail: lc0121@ustc.edu.cn).}
}

\maketitle

\begin{abstract}
Kink effect is a large obstacle for the cryogenic model of inversion-type bulk silicon MOSFET devices. This letter used two methods to correct the kink effect: the modified evolutionary strategy (MES) and dual-model modeling (BSIM3v3 and EKV2.6). Both methods are based on the principle of kink effect. The first method considers impact ionization and substrate current induced body effect (SCBE), and the other considers the change of the freeze-out substrate potential. By applying the above two methods, kink can be corrected to improve the agreement between simulation data and measurement data, and obtain more accurate model parameters. These two methods can be used in further work for cryogenic device modeling and circuit design. 
\end{abstract}


\section{Introduction}
\label{sec:introduction}
\IEEEPARstart{C}{ryogenic} CMOS technology plays an important role in readout and control circuits of quantum chips\cite{b1}\cite{b2}. The accurate device model is a necessary condition for circuits. Due to the huge temperature gap and the kink effect at cryogenic temperatures, industrial models are not suitable for cryogenic circuits simulation. Hence, many studies in recent years have focused on characterization and modeling at cryogenic temperatures\cite{b3,b4,b5,b6,b7}.

Kink effect is common in  cryogenic characteristics of inversion-type bulk silicon MOSFETs, and it is also an obstacle in the modeling process. The double threshold voltage (V$ _{TH} $) model and non-linear resistor have been proposed to correct the kink effect\cite{b8,b9}. 

In this paper, we present MES to implement the correction. Besides, since the BSIM3v3 model\cite{b12} cannot fit the substrate current (I$ _{SUB} $) test results well at liquid helium temperature (4.2K), we use the EKV2.6 model\cite{b13} to simulate I$ _{SUB} $ and propose a dual-model method to correct kink.

\begin{figure}
	\centering
	\includegraphics[width=\linewidth]{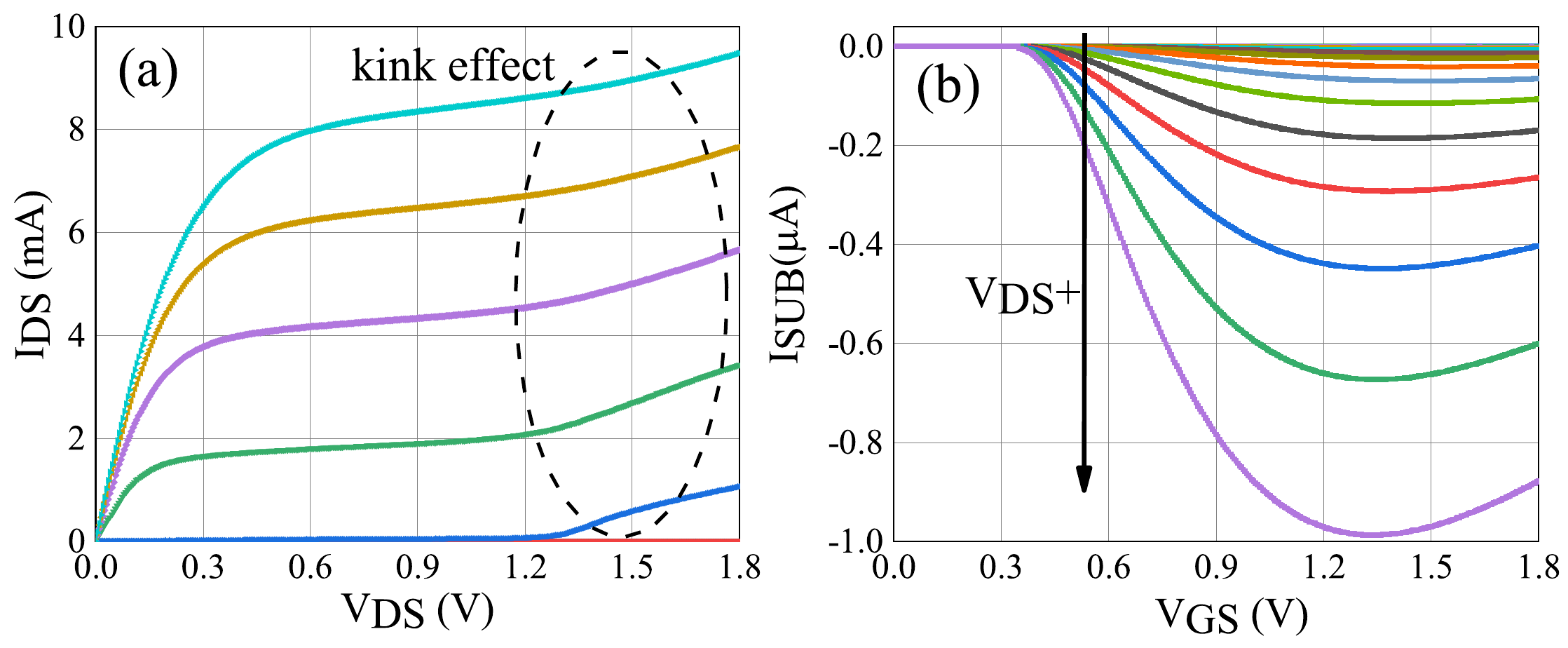}
	\caption{The DUT (NMOS, W/L=10$\mu$m/0.18$\mu$m) I-V characteristics at 4.2K. (a)output characteristics, V$_{DS}$=0V$\rightarrow$1.8V step=5mV, V$_{GS}$=0V$\rightarrow$1.8V step=0.3V, V$_{BS}$=0V; (b)I$ _{SUB} $-V$ _{GS} $, V$_{GS}$=0V$\rightarrow$1.8V$\quad$step=5mV, V$_{DS}$=1V$\rightarrow$ 1.8V step=50mV, V$_{BS}$=0V.}
	\label{fig1}
\end{figure}

\section{Measurement}
The device under test (DUT)  in this study is manufactured with SMIC 0.18$\mu$m bulk CMOS process, that the metal layers in the back end of line (BEOL) are made of aluminum. The DUT is a n-type MOSFET with 1.8V nominal voltage and W/L = 10$\mu$m/0.18$\mu$m. 
We measured the DUT's characteristics at 4.2K, including I$_{DS}$ and I$_{SUB}$, as shown in Fig. \ref{fig1}. All electrical measurements were performed using the Keysight B1500A semiconductor device analyzer.

\section{Simulation Approach and Discussion}

The kink phenomenon is very obvious in Fig. 1(a), and was first explained in \cite{b10}. For DUT (bulk silicon NMOS) at cryogenic temperatures, channel electron can gain enough energy and produce electron-hole pairs by impact ionization when the V$ _{DS} $ is large enough. Electrons flow into the drain, and the corresponding holes migrate towards the substrate (I$ _{SUB} $), resulting in the increase of the substrate potential. Hence V$ _{TH} $ reduces and I$ _{DS} $ increases\cite{b15}\cite{b16}. This is similar to the substrate current induced body effect (SCBE) \cite{b54}, which has been included in the BSIM3v3 model\cite{b12}. Therefore, we try to use the default parameter in BSIM3v3 model to correct the kink effect in Section A. Moreover, in Section B, we considered both carriers impact ionization and the freeze-out substrate, modified the default equation in BSIM3v3, and physically correct the kink effect.

\subsection{MES Modeling}
To obtain an accurate cryogenic model, the MES was proposed to optimize the default parameters of the BSIM3V3 model. The (1+1) evolutionary strategy\cite{b20} was employed in the MES. The first generation parameters' vector was the extracted model parameters in our previous work\cite{b22}. 
And the mutation is the foremost step in MES. 
We selected the optimized parameters of each step by referring to the BSIM3v3 extraction routine, and used MES to adjust these parameters, as shown in Table \ref{tab2}.


One parameter in the  parameters' vector was randomly selected as the mutant gene in each step.  Since the extracted parameters were optimized by people, we adopted the method of adding or subtracting part of the parameters' values to carry out variation instead of the traditional mutation operation to  avoid local optimization. 

Equations of the mutation operation are presented below:
\begin{equation}
\left\{
\begin{array}{l}
i = random.randint(0,26) \\
j = random.randint(-100,100)\\
Gene_i(child) = Gene_i(parent) \times (1+\frac{j}{10000})
\end{array}
\right.
\label{equ1}
\end{equation}
$ Gene_i(child) $ is the value of (i+1)-th parameter of the child generation and $ Gene_i(parent) $ is the value of (i+1)-th parameter of the parent generation. The child parameters' vector is written into the model file after the mutation operation, and the I-V curves and simulation data under the same measure bias are obtained with the simulator Spectre from Cadence.

The measured data characterizes DUT at 4.2K, including transfer characteristics in linear (V$_{DS}$=50mV)  and saturation regions (V$_{DS}$=1.8V),
output characteristics under V$_{BS}$=0V and V$_{BS} $=-1.8V. Since the RMS error is calculated to describe the accuracy of model parameters for every graphics, the fitness function is defined as:

\begin{equation}
\left\{
\begin{array}{l}
RMS \, Error = \sqrt{\frac{1}{N} \times \sum_{i=1}^{n}(\frac{I_{measi}-I_{calci}}{I_{measimax}})^{2} } \times 100\\
\\
fitness = \frac{1}{m} \times \sum\limits_{m} RMS Error,RMS Error
\end{array}
\right.
\label{equ2}
\end{equation}
where N represents the number of data points, $I_{measi}$ represents measured data and $I_{calci}$ is simulated data under the same bias. The value of  $I_{measimax}$  is set to the maximum measured value in the selected graphic according to the BSIMProPlus manual. Letter $ m $ is the number of graphics. For fitness, we refer to the extraction routine to select the corresponding or related RMS Error. According to the fitness value, the program selects the better parameters' vector from the child and parent as the parent of the next generation. Follow the steps to run the optimization program and we get the optimized model file.

\begin{table}
	\caption{Optimization steps and corresponding parameters}
	\centering
	\label{table}
	\setlength{\tabcolsep}{3pt}
	\begin{tabular}{|p{30pt}|p{80pt}|}
		\hline
		Step&Parameters\\
		\hline
		step1& vth0,k1,u0\\
		step2& k2,ua,ub,uc\\
		step3& lint\\
		step4& dvt0,dvt1,dvt2,nlx \\
		step5& voff,nfactor \\
		step6& dwb \\
		step7& vsat,a0,ags \\
		step8& pclm \\
		step9& eta0,etab,dsub \\
		step10& keta \\
		step11& alpha,beta0,alpha1 \\
		step12& all\\
		\hline
	\end{tabular}
	\label{tab2}
\end{table}

\begin{figure}
	\centering
	\includegraphics[width=\linewidth]{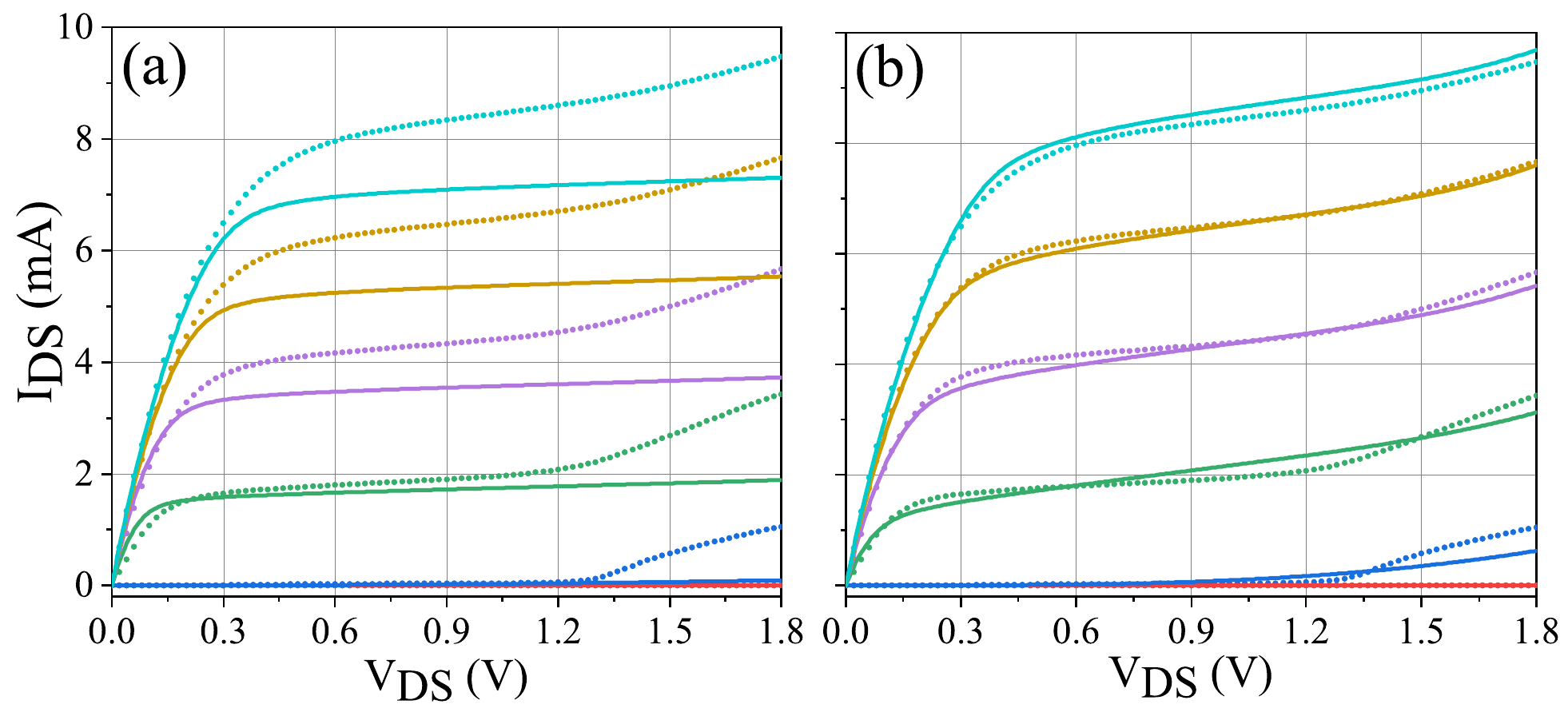}
	\caption{Output characteristics at 4.2K under V$_{BS}$=0V, measurements (scatter plots) and simulations (solid lines) with (a)model in the previous work\cite{b22} and (b)optimized model by MES. }
	\label{fig2}
\end{figure}

The measured output characteristics (V$_{DS}$=0V$\rightarrow$1.2V) are chosen in the program to avoid too small RMS Error. We use the optimized model file for simulation, and the results are shown in Fig. \ref{fig2}(b). Fig. \ref{fig2}(a) displays the simulated output characteristics of the DUT under V$_{BS}$=0V. Although the simulation curve (V$ _{GS} $=0.6/0.9V) still has a certain gap with the measured data in Fig. \ref{fig2}(b), compared with Fig. \ref{fig2}(a), it can be concluded that the accuracy of the optimized model is better than the original model. 

\subsection{Dual-model Modeling}

In Section A, we modified the kink effect by considering carriers impact ionization. However, this method is more like a numercial correction, because the value of modeified I$_{DS}$ is several magnitude lager than the measured I$_{SUB}$.

When the substrate electrode is disconnected from the ground at room temperature, the kink effect also appeared, which is consistent with the result of Hafez $et$ $al.$\cite{b50}. It effectively shows that the kink effect is not only related to impact ionization, but also related to the freeze-out substrate at cryogenic temperature. Due to the freezing of impurities, the substrate is so resistive that it is at floating potential\cite{b51}. The holes produced by impact ionization flow to the freeze-out substrate and accumulate, raising the floating substrate potential, thus resulting in the kink effect\cite{b52}\cite{b53}, as shown in Fig. \ref{fig4}(a). Following the above physical mechanism, the revised threshold voltage expression is:
\begin{equation}
V_{TH_{-} k i n k}=V_{FB}+2\phi_{F}+\gamma \sqrt{2\phi_{F}-V_{BULK}}
\end{equation}

\noindent where $\phi_{F}$ is the Fermi potential, $\gamma$ is the substrate bias effect coefficient, and V$ _{BULK} $ is the potential of the freeze-out substrate, which can be given by a simple expression with: $V_{BULK}=I_{SUB} \times R_{BULK}$, where I$ _{SUB} $ and R$ _{BULK} $ is the substrate current and substrate resistance.

\begin{figure}
	\centering
	\includegraphics[width=\linewidth]{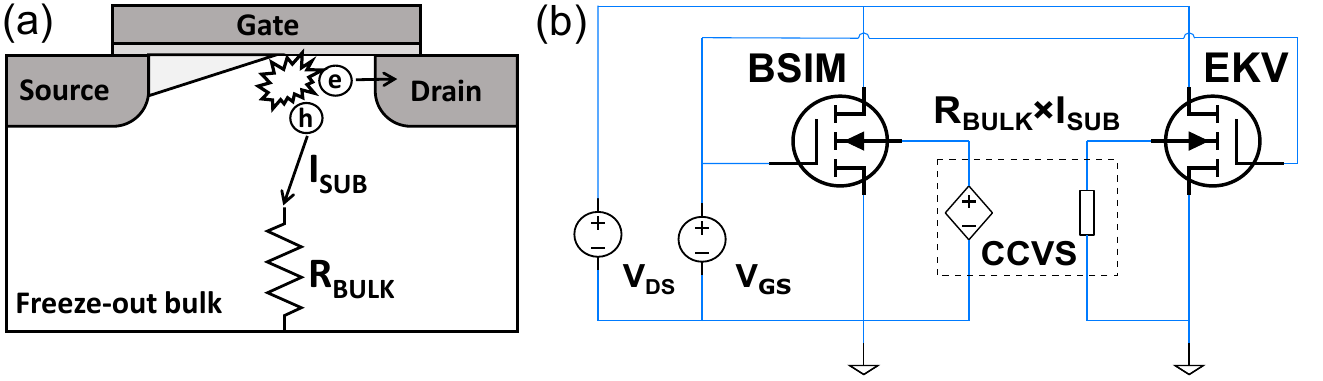}
	\caption{(a)Cross-section of the DUT at 4.2K; (b)the simulation schematic with the dual-model.}
	\label{fig3}
\end{figure}

Our cryogenic MOS model is based on BSIM3v3. However, the BSIM3v3 model cannot match I$ _{SUB} $ test results well at liquid helium temperature. Therefore, we use the EKV2.6 model\cite{b13} to simulate it, as shown in Fig. \ref{fig5}(a). When the V$ _{GS} $ is large enough, the simulation results of I$ _{SUB} $ are consistent with the measurement results, while the simulation results of lower V$ _{GS} $ are slightly larger than the measurement results. This is ascribed to the change of substrate resistance. When I$ _{SUB} $ is small, the substrate is more resistive\cite{b5}, so that some holes can not reach the substrate electrode, but accumulate in the freeze-out substrate. Although there is a certain error between the measurement results and the simulation results, the agreement is better than BSIM3V3. This may be because the reference point of the EKV model is substrate, which has good symmetry so it is more suitable for calculating substrate current. 

Therefore, I$ _{SUB} $ of DUT is calculated by EKV model, then the results are imported into BSIM3v3 model for I$ _{DS} $ calculation. Because the effect of V$ _{BULK} $ and substrate potential on threshold voltage is equivalent, I$ _{SUB} $ generates substrate potential V$ _{BULK} $ through current-controlled voltage source (CCVS, R=750 k$\Omega$), and then connect to the substrate electrode to calculate I$ _{DS} $, as shown in Fig. \ref{fig4}(b). Moreover, the dual-model new model with modified V$ _{TH} $(Eq.(3)) is written in Verilog-A code. Fig. \ref{fig5}(b) shows the calculation result of I$ _{DS} $. In the high gate voltage region, where I$ _{SUB} $ calculations are more accurate, the calculation results and the measured results of I$ _{DS} $ are also in good agreement. RMS errors in figures are shown is Table. \ref{tab3}. It shows that both methods presented in this letter can effectively correct the kink effect. The EKV-BSIM dual model considering the freeze-out substrate potential is based on the physical principle of kink effect, so it has the minimum RMS error.

\begin{figure}
	\centering
	\includegraphics[width=\linewidth]{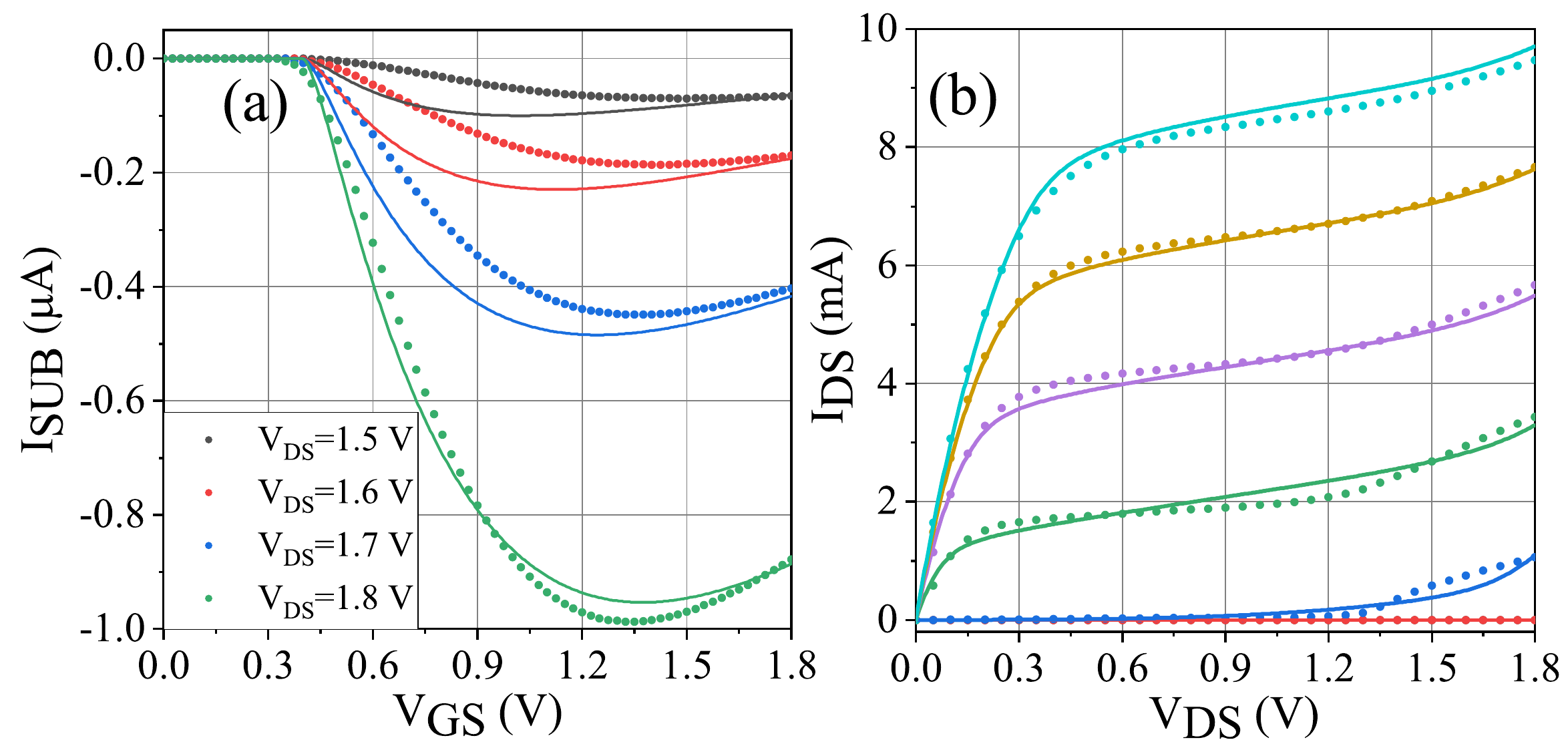}
	\caption{I$ _{SUB} $-V$ _{GS} $ and output characteristics at 4.2K under V$_{BS}$=0V, measurements (scatter plots) and simulations (solid lines). (a)simulation with EKV model; (b)simulation with BSIM3v3. }
	\label{fig4}
\end{figure}
\begin{table}[h]
	\caption{RMS Errors in figures}
	\centering
	\setlength{\tabcolsep}{3pt}
	\begin{tabular}{|p{50pt}|p{50pt}|p{50pt}|}
		\hline
		Fig. \ref{fig2}(a)&Fig. \ref{fig2}(b)&Fig. \ref{fig4}(b)\\
		\hline
		8.41\%& 1.31\%& 1.19\%\\
		\hline
	\end{tabular}
	\label{tab3}
\end{table}

\section{Conclusion}

This work presents two methods to correct the kink effect. The model proposed in Section A does not need to change the original model formula of BSIM3v3, which is more like a numerical model rather than a physical model. The EKV-BSIM dual model proposed in Section B is a compact model based on the physical principle of kink effect, and also verifies the feasibility and superiority of the dual-model modeling. Although many basic formulas of EKV model and BSIM model are similar, they will show their respective advantages in some aspects. In addition, R$ _{BULK} $ is nonlinear due to the varying impurity ionization rate corresponding to electric field. Further work will be carried out to study the varying R$ _{BULK} $ to improve the correction of kink effect.





\end{document}